# Radio-frequency pulse design in local rotating frame in magnetic resonance imaging

Seung-Kyun Lee*

GE HealthCare Technology and Innovation Center, Niskayuna, NY 12309, USA

## Abstract

The problem of spatially selective radio-frequency (RF) pulse design in magnetic resonance imaging (MRI) is typically stated in the form of determining, analytically or numerically, RF waveforms to be applied in synchrony with one or more predetermined gradient waveforms. In most cases, the dynamics of the nuclear spin magnetization under the RF and gradient fields is described in a global rotating frame that cancels the effect of the static (main) magnetic field $B_0$. In this work, we consider an alternative frame of reference, which can be called a local rotating frame where total longitudinal magnetic field ($B_0$ plus gradient) in every voxel is zero. In this frame, the effect of time-dependent gradient field is integrated out, and the remaining magnetization dynamics, governed by much weaker RF fields, becomes both simpler and slower. We show that recasting existing RF design methods in such a frame provides useful insights and techniques that are not obvious in the conventional description. The methods we consider include (i) two-dimensional spatial RF pulse design in the excitation k-space, (ii) Shinnar-Le Roux RF pulse design, (iii) residual phase calculation in slice-selective excitation, and (iv) iterative and numerical solutions for multi-coil RF pulse design. In particular, we show that the new formalism can substantially reduce the Bloch simulation time which can greatly benefit iterative pulse design in parallel transmit. In all, the proposed framework provides considerable theoretical insights and practical utility for RF pulse design in MRI.

*Email: lsk@gehealthcare.com

Word Count: ~6000 (main text), 400 (Appendix)

## 1. Introduction

A spatial radio frequency (RF) pulse in MRI can be defined as a pulse that creates spin excitation with spatial profile different from that of the RF field itself. Such spatial "tailoring" is achieved by a gradient pulse that is played simultaneously with the RF pulse. Whereas a classic example of a spatial RF pulse is a one-dimensional, slice-selective pulse with a single gradient and an RF field source, interest in multidimensional, multi-coil RF design has grown rapidly in recent years as such design was used to mitigate image shading caused by *in vivo* RF field distortion in modern high-field MRI. In a multi-coil RF transmission scheme, known as parallel transmit [1], [2], transmit RF field ($B_1^+$) profile of each coil is first determined by either amplitude- or phase-based $B_1^+$ mapping method *in vivo*. The RF waveforms of individual coils are then determined based on these maps with a goal of achieving the best excitation accuracy subject to RF power limitation.

In most parallel transmit RF design methods, gradient waveforms are determined separately from, and prior to, the RF waveform calculation. Hardware constraints and the spatial frequency content of the target magnetization and $B_1^+$ field profiles usually dictate the choice of the gradient waveforms, and the primary task of spatial RF design often boils down to optimizing RF waveforms for predetermined gradient pulse shapes. Several authors have explored joint RF and gradient design. For example, Yip et al [3] proposed a method in which gradient-dependent RF optimization alternates with parametric adjustment of echo-planar type gradient trajectory. Sparse pulse design [4] is another example which involves active gradient optimization. In these methods, gradient-specific RF calculation is still important as a sub-problem in a multi-stage optimization process.

In this paper we focus specifically on the problem of RF calculation for a predetermined gradient pulse. We show that such a problem is significantly simplified when one takes advantage of the prior knowledge of the gradient waveform and transforms away the gradient field through a novel rotating frame transformation. In contrast to the conventional rotating frame which rotates at a fixed Larmor frequency of the static field $B_0$, the proposed frame rotates at a position- and time- dependent angular rate determined by the total longitudinal field, including $B_0$, gradient field, and any known (static) variation in $B_0$. In this approach, RF pulse design is formulated in a frame, which may be called a "local rotating frame", where only transverse magnetic fields are present. The frame transformation is well defined for a given gradient pulse and known static field variation so that RF waveforms obtained in one frame can be translated to the other in a straightforward manner. An immediate consequence of the proposed transformation is that in the new frame, apparent RF field rotates at a voxel dependent rate, even if the applied field is spatially homogeneous in the laboratory frame. As an illustration, in the local rotating frame, a slice selective RF pulse does not excite out-of-slice spins because the RF field rotates too fast in those locations. Although this fact may seem complicating, we show below that, in case of Bloch simulation, additional complexity due to apparent RF rotation is well outweighed by the benefit of slower spin dynamics provided by the local rotating frame.

We note that other researchers have also proposed voxel-dependent frame transformation in Bloch simulation [5] and selective pulse design [6]. Our method is different in that only the longitudinal field, which is predetermined and easily integrable, is transformed away. Unlike in [6], the transformation is not tied to a particular pulse design algorithm, but rather is introduced to support and simplify RF design methods in its many varieties including those for parallel transmit.

It is the goal of this paper to explore the potential benefits of the new rotating frame approach in spatial RF pulse design. Whereas the method is general, we put particular emphasis on design schemes applicable to multidimensional pulses in parallel transmit. In the theory section, we define the transformation and discuss the solution of the Bloch equation formulated in the new frame. We take two examples in one-dimensional pulse design to demonstrate the utility of the new formulation. Next we extend the method to parallel transmit and demonstrate a phase-constrained, iterative RF design scheme analogous to the one in [7]. Finally we explain how the proposed frame transformation helps speed up Bloch simulation and demonstrate reduction in optimization time in optimal control-based iterative pulse design [8].

## 2. Theory

### 2.1. Transformation in the magnetization domain

Conventionally, MRI pulse design is formulated in a frame rotating at a fixed Larmor frequency $\omega_0 = -\gamma B_0$. Here $\gamma$ is the gyromagnetic ratio, positive for proton. We will use subscript $c$ to denote variables referenced to the conventional frame. For example, longitudinal and transverse magnetization in the conventional frame at position $\vec{r}$ and time $t$ are denoted as $M_{zc}(\vec{r},t)$ and $M_{xyc}(\vec{r},t) \equiv M_{xc} + iM_{yc}$, respectively. We define a "local rotating frame" as a frame which has a transverse phase $-\phi(\vec{r},t)$ with respect to the conventional frame, where $-\phi$ is the integral of the instantaneous Larmor frequency in the conventional frame,

$$\begin{aligned} -\phi(\vec{r},t) &\equiv -\gamma \int_0^t B_z dt' \\ &= -\gamma \int_0^t \left(\delta B_0(\vec{r}) + \vec{G}(t') \cdot \vec{r}\right) dt' \end{aligned} \tag{1}$$

determined by the static off-resonance field $\delta B_0(\vec{r})$ and the applied gradient field $\vec{G}(t)$. The magnetization components in the local rotating frame, $M_{xy} \equiv M_x + iM_y$ and $M_z$, are related to those in the conventional frame as

$$\begin{aligned} M_{xy} &= M_{xyc} e^{i\phi(\vec{r},t)} \\ M_z &= M_{zc}. \end{aligned} \tag{2}$$

The corresponding polar ($\theta_l$) and azimuthal ($\phi_l$) angles of magnetization in the local rotating frame are

$$\begin{aligned} \theta_l &= \theta_c \equiv \theta \\ \phi_l &= \phi_c + \phi(\vec{r},t). \end{aligned} \tag{3}$$

Likewise, the complex transverse RF magnetic field $B$ in the local rotating frame is given by

$$B(\vec{r},t) \equiv B_x + iB_y = B_c(\vec{r},t) e^{i\phi(\vec{r},t)} \tag{4}$$

where $B_c(\vec{r},t)$ is the complex RF field in the conventional frame. For non-interacting spins the above transformation effectively eliminates longitudinal fields from the Bloch equation, which now takes a simpler form:

$$\frac{d}{dt}\begin{pmatrix} M_x \\ M_y \\ M_z \end{pmatrix} = \gamma \begin{pmatrix} 0 & 0 & -B_y \\ 0 & 0 & B_x \\ B_y & -B_x & 0 \end{pmatrix} \begin{pmatrix} M_x \\ M_y \\ M_z \end{pmatrix}. \tag{5}$$

Here and in what follows we will ignore relaxation, concerning ourselves only with RF pulses which are short compared to the relaxation times. Also we assume that the magnetization vector is normalized to unity, $|M_{xy}|^2 + |M_z|^2 = 1$.

### 2.2. Riccati form

Eq. (5) can be further simplified by mapping the magnetization vector onto a single complex variable $f$ defined as

$$f \equiv 2\frac{M_x + iM_y}{1 + M_z}. \tag{6}$$

This mapping converts Eq. (5) into a so-called Riccati form:

$$\frac{df}{dt} = i\gamma B(\vec{r},t) - i\gamma B^*(\vec{r},t)\frac{f^2}{4}. \tag{7}$$

The initial condition is $f(t=0)=0$ for magnetization originally aligned along the $z$-axis.

The Bloch-Riccati equation and the mapping Eq. (6) have been extensively used in the literature for one-dimensional selective RF design; examples include the hyperbolic secant pulse [9], [10], stereographic projection method [11], and the inverse scattering algorithm [12], [13], [14], [15], [16], [17]. Among different techniques to solve Eq. (7), the one that is most amenable to extension to multidimensional, multi-coil design is iterative inversion [9], [12]. In this section we derive from Eq. (7) an integral representation of the forward solution of the Bloch equation; we demonstrate iterative backward solution in section 4.

First, we integrate Eq. (7) from $t=0$ to $T$,

$$f(T) = i\gamma\int_0^T B(\vec{r},t)dt - i\gamma\int_0^T B^*(\vec{r},t)\frac{f^2(t)}{4}dt\ . \tag{8}$$

We now define a quantity

$$q \equiv \frac{2\tan^{-1}|f/2|}{|f|}f \tag{9}$$

which is a complex number with amplitude $|q|=\theta$, and phase $\arg(q)=\phi_l$. Both $f$ and $q$ can be viewed as a scaled transverse magnetization; see Fig. (1a) for pictorial definition of $f$ and $q$. Up to $\theta \approx 90°$, $q$ follows $f$ closely. Indeed, for $|f| \leq 2$, Eq. (9) Taylor-expands as

$$q = \left(1 - \frac{1}{3}\left|\frac{f}{2}\right|^2 + \frac{1}{5}\left|\frac{f}{2}\right|^4 + \cdots\right)f. \tag{10}$$

Combining Eqs. (8, 10), we obtain a formal, series expansion solution of the Bloch equation in $q$, with terms arranged in an increasing order of $f$:

$$q(T) = i\gamma\int_0^T B(\vec{r},t)dt - i\gamma\int_0^T B^*(\vec{r},t)\frac{f^2(t)}{4}dt - \frac{1}{3}\left|\frac{f(T)}{2}\right|^2 i\gamma\int_0^T B(\vec{r},t)dt + O((f/2)^4). \tag{11}$$

The first term dominates the series if $|f(t)/2|^2 \ll 1$ for all $t$, $0 \leq t \leq T$. Figure (1b) indicates that this is likely satisfied for $\theta(T)$ up to ~70°. However, we now present an argument that in fact, the first term is likely to lead the entire series up to $\theta(T)=90°$.

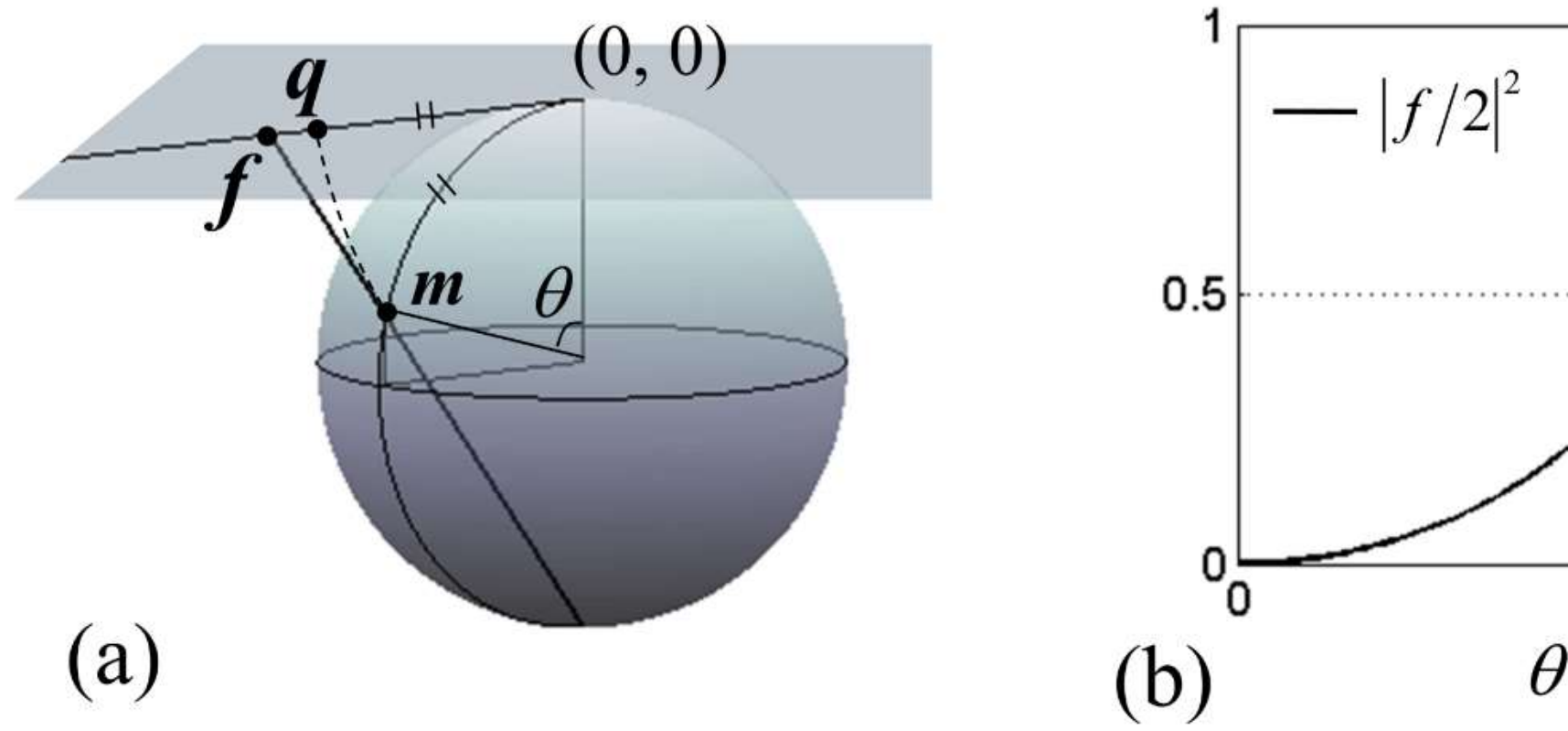

**Figure 1**. (a) Graphical definition of the scaled complex magnetization $f$ and $q$. $f$ is called stereographic projection. (b) Dependence of the magnitude of $f$ on the polar angle $\theta$ of magnetization.

First, the third and the following terms in Eq. (11) represent the difference between $f$ and $q$. From Eq. (9), the fractional difference between $f$ and $q$ is such that the first two terms in Eq. (11) dominate the rest of the series by a ratio $1{:}\,1-\frac{2\tan^{-1}|f/2|}{|f|}$. This is $1{:}\,0.215$ when $|f(T)/2|=1$. Therefore even at $\theta(T)=90°$ the first two terms still account for nearly 80% of the amplitude of the right-hand side of Eq. (11). Next we compare the first and second terms. We note that the second integral in the series of Eq. (11) is an integral of $B^*$ weighted by $f^2(t)/4$. During the course of typical excitation $\theta$ grows roughly monotonically. Therefore, for a 90° pulse, $|\,f^2(t)/4\,|$ is likely substantially smaller than unity for most of the time. This means that the weighted sum (the second integral) should be significantly smaller than the direct sum (the first integral) of $B^*(\vec{r},t)$, unless the phase of $B(\vec{r},t)$ varies much faster than $B^*(\vec{r},t)\,f^2/4$. The latter situation is unlikely because both $B(\vec{r},t)$ and $B^*(\vec{r},t)\,f^2/4$ contain a single power of $e^{i\phi(\vec{r},t)}$ and a single power of $B_c(\vec{r},t)$; whether a particular voxel is dominated by gradient or RF field, the two integrands must advance in phase at a comparable rate. We conclude, therefore, that the second term in the right-hand side of Eq. (11) is likely substantially smaller than the first even if at the end of the pulse $|f(T)/2|=1$.

Numerical experiments show that the expansion Eq. (11) is valid in the sense that iterative inversion based on the first term converges. This is so under a wide range of experimental conditions in which $\theta(T)\le 90°$. We show examples in section 4.

### 2.3. Small-tip-angle design, revisited

Retaining only the first term in the right-hand side of Eq. (11) gives the defining equation for the linear (Fourier) pulse design. Starting from the approximation,

$$q(T)\approx i\gamma\int_0^T B(\vec{r},t)\,dt \tag{12}$$

we rewrite Eq. (12) back in the conventional frame by substituting Eq. (4) for $B$ and Eq. (1) for $\phi$. The scaled complex magnetization at the end of the pulse in the conventional frame is then given by

$$\begin{aligned} q_c(T) &\equiv \theta(T)e^{i\phi_c(T)} \\ &= q(T)\,e^{-i\phi(\vec{r},T)} \end{aligned}$$

$$\approx i\gamma \int_0^T B_c(\vec{r},t)\, e^{i\phi(\vec{r},t)-i\phi(\vec{r},T)}\, dt$$

$$= i\gamma \int_0^T B_c(\vec{r},t) \exp\left(-i\gamma \int_t^T \vec{G}\cdot\vec{r}dt' + i\gamma(t-T)\delta B_0(\vec{r})\right) dt \tag{13}$$

We identify the excitation *k*-space vector $-\gamma \int_t^T \vec{G}dt'$ [18] and off-resonance correction [19] in the integration kernel, which all came about naturally in the process of frame transformation. Eq. (13) is different from small-tip-angle design equations used by other authors, e.g., [19], [20], [21], in an important way; we are dealing with the scaled magnetization $q_c = \theta e^{i\phi_c}$, not transverse magnetization. In the literature, Eq. (13), with $q_c$ replaced by $M_{xyc}$, is widely used for arbitrary tip-angle pulse design [2], [7], [21], [22] under the condition that (i) off-resonance field is negligible, (ii) gradient trajectory belongs to a certain class, and (iii) the RF field $B_c$ at every spatial point is Hermitian symmetric with respect to such trajectory. When any of these conditions is violated, the Linear Class Large Tip Angle (LCLTA) pulse theory [22] does not explain why Eq. (13) would work, as it often does experimentally, at tip angles much in excess of 30°. What we have shown here is that for an intermediate tip angle, up to 90°, Eq. (13) is a valid leading-order approximation of the series solution Eq. (11) without any of the above conditions, and therefore is expected not to incur excessive error when used in the presence of off-resonance field, with an arbitrary gradient trajectory, or without particular symmetry of the RF pulse.

Before we demonstrate applications of Eqs. (11, 13) in parallel transmit, we complete the theory section with the local rotating frame transformation in the spinor domain. This will lead us to another widely used method of Shinnar-Le Roux (SLR) pulse design.

## 2.4. Transformation in the spinor domain

RF pulse design theory can be formulated in the spinor domain [22], [23], [24], [25]. Here the Bloch equation governs the time evolution of the complex rotation parameters $\alpha(t)$ and $\beta(t)$, called the Cayley-Klein parameters, and the magnetization at $t = T$ is obtained by applying the rotation operator at $t = T$ to magnetization at $t = 0$. Since knowledge of $(\alpha, \beta)$ permits knowledge of magnetization, but not vice versa, pulse design in $(\alpha, \beta)$ is more versatile. For example, a refocusing pulse can be designed unambiguously in terms of $(\alpha, \beta)$. Here we use the convention of [22], [23], [26] to define $\alpha$, $\beta$ in terms of real-space rotation. Specifically, rotation in three dimensions around a unit vector $(n_x, n_y, n_z)$ by an angle $\epsilon$ (right hand rule) is represented by a matrix

$$Q = \begin{pmatrix} \alpha & \beta \\ -\beta^* & \alpha^* \end{pmatrix} \tag{14}$$

with prescription

$$\alpha = \cos\frac{\epsilon}{2} - in_z \sin\frac{\epsilon}{2},\ \ \beta = -i(n_x - in_y)\sin\frac{\epsilon}{2}. \tag{15}$$

Successive physical rotations $Q_1$, $Q_2$, ..., $Q_n$ map onto a product matrix $Q_n \cdots Q_2 Q_1$.

Local rotating frame transformation of $(\alpha, \beta)$ can be obtained along the following line. Let us denote the Cayley-Klein parameters in the conventional frame as $\alpha_c$ and $\beta_c$. These represent particular physical rotation of magnetization that occurred at position $\vec{r}$ during time $0 \sim t$. How does the same physical rotation appear in a frame which has gained transverse phase $-\phi(\vec{r},t)$ with respect to the original frame? The apparent rotation is constructed in two steps. First the initial magnetization counter-rotates, by $\phi(\vec{r},t)$, around the $z$ axis. Second, this new magnetization undergoes rotation defined by the Cayley-Klein parameters $(\alpha_c, \beta_c e^{-i\phi})$. The additional factor $e^{-i\phi}$ accounts for the azimuthal shift in the magnetization rotation axis when viewed in the new frame. The result is

$$\alpha(\vec{r},t) = \alpha_c(\vec{r},t)e^{-i\phi(\vec{r},t)/2}$$
$$\beta(\vec{r},t) = \beta_c(\vec{r},t)e^{-i\phi(\vec{r},t)/2}. \quad (16)$$

If we substitute Eqs. (1,4,16) to the Bloch equation satisfied by the conventional-frame quantities $\alpha_c, \beta_c, B_c, B_z$ (see, for example, ref. [22]), that is,

$$\frac{d}{dt}\alpha_c^* = \frac{i\gamma}{2}(B_c(\vec{r},t)\beta_c - B_z\alpha_c^*)$$
$$\frac{d}{dt}\beta_c = \frac{i\gamma}{2}(B_z\beta_c + B_c^*(\vec{r},t)\alpha_c^*)$$
$$\alpha_c(0) = 1$$
$$\beta_c(0) = 0, \quad (17)$$

we obtain the Bloch equation in the new frame

$$\frac{d}{dt}\alpha^* = \frac{i\gamma}{2}B(\vec{r},t)\beta$$
$$\frac{d}{dt}\beta = \frac{i\gamma}{2}B^*(\vec{r},t)\alpha^*$$
$$\alpha(0) = 1$$
$$\beta(0) = 0. \quad (18)$$

As expected, Eq. (18) involves no longitudinal field. Various attempts can be made to solve Eq. (18). For example, Eq. (18) reduces to the Riccati form Eq. (7) upon definition $f \equiv -2\beta^*/\alpha$. In the following section, we present a finite-difference time-domain approach and a series expansion solution of Eq. (18) applied to one-dimensional selective RF pulse calculation.

## 3. Applications to one-dimensional selective pulse problem

### 3.1. Alternative derivation of the forward SLR transform

Eq. (18) is exactly integrable if $B$ is constant in time. While this cannot be in the local rotating frame, an approximate solution can be constructed if $B$ is approximated by a piecewise constant function on a regular time grid,

$$B(\vec{r},t) \to B_n \equiv B(\vec{r},t_n)$$
$$t_n = n\Delta t, n = 1,2,\cdots \quad (19)$$

with a finite time step $\Delta t$. Starting from the initial condition $\alpha_0 = 1, \beta_0 = 0$, Eq. (18) can be solved for $\alpha_n \equiv \alpha(\vec{r},t_n), \beta_n \equiv \beta(\vec{r},t_n)$ recursively,

$$\begin{pmatrix}\alpha_n^* \\ \beta_n\end{pmatrix} = \begin{pmatrix}C_n & S_n \\ -S_n^* & C_n\end{pmatrix}\begin{pmatrix}\alpha_{n-1}^* \\ \beta_{n-1}\end{pmatrix}, n = 1,2,\cdots \quad (20)$$

where

$$C_n \equiv \cos(\gamma|B_n|\Delta t/2)$$
$$S_n \equiv i\frac{B_n}{|B_n|}\sin(\gamma|B_n|\Delta t/2). \quad (21)$$

The first few solutions are

$$\begin{pmatrix}\alpha_1^* \\ \beta_1\end{pmatrix} = \begin{pmatrix}C_1 \\ -S_1^*\end{pmatrix}$$

$$\begin{pmatrix} \alpha_2^* \\ \beta_2 \end{pmatrix} = \begin{pmatrix} C_2C_1 - S_2S_1^* \\ -S_2^*C_1 - C_2S_1^* \end{pmatrix}$$

$$\begin{pmatrix} \alpha_3^* \\ \beta_3 \end{pmatrix} = \begin{pmatrix} C_3C_2C_1 - C_3S_2S_1^* - S_3S_2^*C_1 - S_3C_2S_1^* \\ -S_3^*C_2C_1 + S_3^*S_2S_1^* - C_3S_2^*C_1 - C_3C_2S_1^* \end{pmatrix}. \tag{22}$$

These expressions appear more illuminating if we restrict ourselves to one-dimensional RF pulse with a constant gradient field $\vec{G}(t) = \vec{G}$. In such a case, the discretized RF field in the local rotating frame is related to that in the conventional frame by (see Eq. (4))

$$B_n = B_{nc}\rho^n \tag{23}$$

where

$$\rho \equiv \exp\left(i\gamma\left(\vec{G} \cdot \vec{r} + \delta B_0(\vec{r})\right)\Delta t\right) \tag{24}$$

represents $\vec{r}$-dependent phase advance in $\Delta t$. Upon substituting Eq. (23) to Eqs. (21, 22), we find that at every time step, the Cayley-Klein parameters reduce to two complex polynomials in $\rho^{-1}$,

$$\alpha_n(\vec{r}) = \sum_{j=0}^{n-1} a_j\rho^{-j}$$

$$\beta_n(\vec{r}) = \sum_{j=1}^{n} b_j\rho^{-j} \tag{25}$$

with coefficients $a_j, b_j$ determined by the RF field $B_{jc}, j = 1, 2, \ldots, n$ in the conventional frame. The mapping of $B_{jc}$ onto the polynomials in Eq. (25) is known as the forward Shinnar-Le-Roux (SLR) transformation [27], [28]. This is useful because if $B_{jc}$ has no position dependence in the conventional frame, as is often assumed in single-coil selective excitation, Eq. (25) can be inverted by first expressing the desired excitation profile $\alpha_n(\vec{r}), \beta_n(\vec{r})$ as a linear combination of powers of $\rho^{-1}$, and then calculating back the time series $B_{jc}, j = 1, 2, \ldots, n$ from the equal number of coefficient pairs $(a_{j-1}, b_j), j = 1, 2, \ldots, n$.

Conventionally, the SLR transform is derived using "hard pulse approximation" as a key step. This is necessary to separate the effects of the gradient and the RF fields. In the local rotating frame approach, the separation is already *exactly* achieved by frame transformation, so the polynomial expansion of $\alpha$ and $\beta$ does not require hard pulse approximation. In fact, a hard pulse, consisting of an infinitely sharp RF burst followed by a zero-RF interval, can be replaced by a piecewise constant RF field (corresponding to the average of the RF burst over the interval) in the local rotating frame on a regular time grid. Therefore, the hard pulse approximation in the SLR transform can be replaced by a weaker assumption of piecewise constant RF field in the local rotating frame.

### 3.2. Second order phase in slice selective excitation

A traditional slice selective excitation pulse exhibits the effect of the nonlinearity of the Bloch equation at a large tip angle. We have seen in section 2 that at 90°, nonlinear effect may account for about 20% of the linear response. Experimentally, such effect manifests itself as extra phase winding along the slice direction in the excited magnetization [9]. The residual phase is usually removed by empirically adjusting the area of the refocusing gradient after the RF. Here we present analytical calculation of the residual phase in one-dimensional selective excitation. We show that a second-order calculation of Eq. (18) gives a satisfactory result in good agreement with Bloch simulation up to a tip angle of 90°.

Analytical integration of Eq. (18) from $t = 0$ to $T$ produces a series expansion solution for $\alpha(T)$ and $\beta(T)$ in increasing orders of the RF field strength. The small parameter of expansion can be defined as $\Delta = \left|\int_0^T \gamma B dt /2\right|$. This is one half of the tip angle in radians at the center of the gradient in slice selective excitation. Since the successive terms in the series involve integral of increasing powers of $B(t)$, series convergence depends on the peaks of $B(t)$, and therefore is not guaranteed by $\Delta < 1$. Numerical survey indicates, however, the series converges rapidly for most realistic RF waveforms with $\Delta < 1$.

The first few terms in the series solution are shown below. The numbers in the parentheses in the superscript indicate the order of each term in the solution. $\alpha(T)$ contains only even order terms, whereas $\beta(T)$ contains odd:

$$\alpha^*(T) = \alpha^{*(0)}(T) + \alpha^{*(2)}(T) + \alpha^{*(4)}(T) + \cdots .$$
$$\beta(T) = \beta^{(1)}(T) + \beta^{(3)}(T) + \beta^{(5)}(T) + \cdots . \tag{26}$$

where

$$\alpha^{*(0)}(T) = 1$$
$$\beta^{(1)}(T) = \frac{i\gamma}{2}\int_0^T B^*(\vec{r},t)dt$$
$$\alpha^{*(2)}(T) = \left(\frac{i\gamma}{2}\right)^2 \int_0^T \int_0^t B(\vec{r},t)B^*(\vec{r},t')dt'dt$$
$$\beta^{(3)}(T) = \left(\frac{i\gamma}{2}\right)^3 \int_0^T \int_0^t \int_0^{t'} B^*(\vec{r},t)B(\vec{r},t')B^*(\vec{r},t'')dt''dt'dt\,. \tag{27}$$

The first two terms determine the linear response in the spin dynamics. At each order, $\alpha$ and $\beta$ transform to conventional frame quantities according to Eq. (16). As we restrict ourselves to a one-dimensional slice selective pulse with constant gradient $\vec{G} = G\hat{z}$, we use $\phi(\vec{r},t) = \gamma Gzt$, ignoring off-resonance field. The RF field in the conventional frame $B_c(t)$ is assumed to be spatially uniform and as a function of time Hermitian symmetric with respect to $t = T/2$ [22]: $B_c(t) = B_c^*(T-t)$. This includes a conventional sinc pulse. Substituting $B(\vec{r},t) = B_c(t)e^{i\gamma Gzt}$ in the second line of Eq. (27) gives the relationship

$$\beta^{(1)}(t) = ie^{-i\gamma GzT/2} \cdot \text{(real number)}. \tag{28}$$

which reproduces the "linear phase" of $\beta^{(1)}(t)$ valid at small tip angles.

Calculation of the next order term $\alpha^{(2)}(T)$ is shown in the Appendix. Key results are that in one spatial dimension, both the real and imaginary parts of $\alpha^{(2)}(T)$ are completely determined by $|\beta^{(1)}(T)|$. Explicitly,

$$\alpha_{\text{real}}^{(2)}(z,T) \equiv \text{Re}\left(\alpha^{(2)}(z,T)\right) = -\frac{1}{2}\left|\beta^{(1)}(z,T)\right|^2 \tag{29}$$

and

$$\alpha_{\text{imag}}^{(2)}(z,T) \equiv \text{Im}\left(\alpha^{(2)}(z,T)\right) = -\frac{1}{\pi}\int_{-\infty}^{\infty} \frac{\alpha_{\text{real}}^{(2)}(z',T)}{z-z'}dz'\,. \tag{30}$$

With these results we can calculate the second-order phase in slice selective excitation. First, the transverse magnetization is expressed in the Cayley-Klein parameters as [22]

$$M_{xy} = -2\alpha^*\beta^*. \tag{31}$$

At $t = T$, the phase of this quantity in the conventional frame is, to the second order in $\Delta$,

$$\arg M_{xyc} = \arg\left(M_{xy}e^{-i\gamma GzT}\right)$$

$$
\begin{aligned}
&= \arg(-2\alpha^*\beta^*) - \gamma GzT \\
&\approx \arg\left(\alpha^{*(0)} + \alpha^{*(2)}\right) + \arg\left(-\beta^{*(1)}\right) - \gamma GzT \\
&= \arg\left(1 + \alpha_{\text{real}}^{(2)} - i\alpha_{\text{imag}}^{(2)}\right) \pm \pi/2 - \gamma GzT/2\,. \qquad (32)
\end{aligned}
$$

All the magnetization and spinor variables in the above are calculated at position $z$ and time $T$. The term $\gamma GzT/2$ in the last line represents the linear phase that is removed by the usual negative half-area gradient. We define the *residual* phase as $\phi_{res} \equiv \arg M_{xyc} \ + \gamma GzT/2$ which has $z$ dependence due to nonlinearity of the Bloch equation. The slope of $\phi_{res}$ at the center of the slice is

$$
\left.\frac{d\phi_{res}}{dz}\right|_{z=0} = -\frac{d}{dz}\left.\left(\frac{\alpha_{\text{imag}}^{(2)}}{1+\alpha_{\text{real}}^{(2)}}\right)\right|_{z=0} = -\frac{1}{2\pi z_{\text{FW,eff}}}\cdot\frac{\theta_t^2}{1-\theta_t^2/8}. \qquad (33)
$$

The Appendix contains the details of this calculation. Here $\theta_t = 2\left|\beta^{(1)}(0,T)\right| = \left|\int_0^T \gamma B(0,t)dt\right|$ is the tip angle at the center of the slice, and $z_{\text{FW,eff}}$ is an effective slice thickness whose exact value depends on the sharpness of the slice profile; for a rectangular slice profile, it equals the actual thickness (full width) $z_{FW}$. For a 90° pulse with a rectangular slice profile, Eq. (33) gives the phase slope at the center of the slice $d\phi_{res}/dz = -32.5°/z_{\text{FW}}$.

In Fig. 2, we test the validity of Eq. (33). Two different RF pulses, namely Hamming-windowed sinc pulses applied along the $x$-axis with time-bandwidth product $p$ = 12 (Fig. 2a) and 4 (Fig. 2d), were used for the study. The gradient amplitude was scaled to produce nominal slice thickness of 1 cm (along the $z$-axis), and a negative half-area rewinder was used. Explicit Bloch simulation shows magnetization profiles (Fig. 2b,e) with usual dispersion-like behavior in the quadrature component $M_{xc}$. Residual phase slope was obtained by linear fit to the simulated profiles for a given tip angle; the simulation was repeated for tip angles (varied by scaling RF amplitude) of 10° to 90°, and the results were plotted in Fig. (2c,f). In both RF pulses, the simulated data closely match the analytical calculation (solid line). We note that for $p$ = 12, the effective thickness is nearly the same as the nominal thickness, since the slice boundary is relatively sharp. In contrast, for $p$ = 4, broad transition region in slice profile results in $z_{\text{FW,eff}}$ about 30% smaller than 1 cm, and therefore a larger $d\phi_{res}/dz$ according to Eq. (33) which is consistent with the Bloch simulation. In all cases, Eq. (33) accurately predicts residual dephasing for a range of tip angles studied, where the second-order series solution is sufficient.

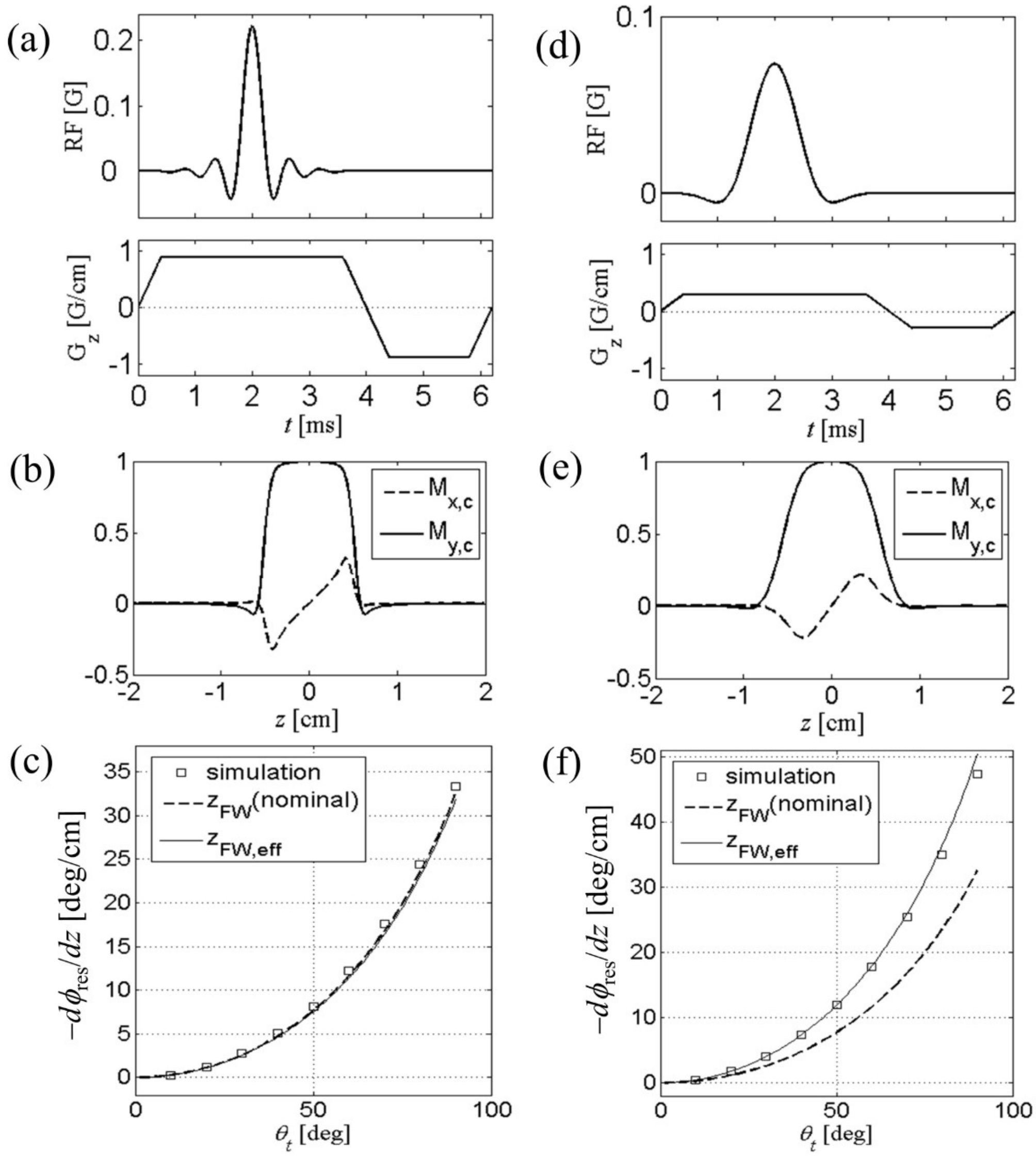

**Figure 2**. (a) RF and gradient waveforms for slice selective 90° excitation. The RF pulse is conventional Hamming-windowed sinc pulse with time-bandwidth product $p$ = 12, and active pulse length of 3.2 ms. The negative "rewinder" portion of the gradient pulse has half the area of the positive portion. (b) Bloch-simulated slice excitation profile. Residual phase winding is apparent from the dispersion-like shape of $M_{xc}$. Subscript $c$ denotes quantities in conventional rotating frame. (c) The phase slope at the center of the slice, as a function of the tip angle. This nonlinear effect is well predicted by analytical calculation Eq. (33). For a large $p$, $z_{\mathrm{FW,eff}}$ is close to the nominal slice thickness. (d-f) Same as in (a-c), except that the RF pulse has $p$ = 4 and the gradient was scaled to give the same nominal thickness of 1 cm.

## 4. Applications to parallel excitation with multiple transmit coils

Explicit nonlinear calculations shown in section 3 cannot be easily extended to a parallel transmit situation. However, iterative inversion of Eq. (11) and fast numerical integration of Eq. (18) apply to any number of transmit coils, which are the subjects of this section.

### 4.1. Iterative inversion

Iterative inversion of the Bloch equation [9], [12], [29] consists of repeated applications of linear inversion followed by Bloch simulation and error calculation. In applying this method to solve Eq. (11), the convergence of the solution depends on (i) the goodness of the linear approximation, Eq. (13), and (ii) how well the linear equation can be inverted. We expect, based on the argument in section 2.2, that the goodness condition is met for tip angles up to 90°. The invertibility of the linear equation depends on the choice of a gradient trajectory. Since Eq. (13) is not subject to the conditions of the LCLTA theory, gradient trajectory is not limited to a particular class. In parallel excitation with linearly independent $B_1^+$ profiles, the gradient pulse can be shortened by transmit sensitivity encoding [2]. Below we demonstrate calculation of an eight-channel parallel transmit RF pulse based on an accelerated echo-planar gradient trajectory.

In parallel transmit, the RF field $B_c$ in Eq. (13) is a weighted sum of the RF waveforms $\{b_1, b_2, \ldots\}$ of $N$ coils,

$$B_c(\vec{r}, t) = s_1(\vec{r}) b_1(t) + s_2(\vec{r}) b_2(t) + \cdots + s_N(\vec{r}) b_N(t) \tag{34}$$

where $\{s_1, s_2, \ldots\}$ are complex $B_1^+$ sensitivity profiles. For a given target excitation profile, Eqs. (13,34) can be solved for $\{b_1, b_2, \ldots\}$ in the least square sense. The procedure for linear inversion is detailed in [2], [20]. For tip angles approaching 90°, the first order solution $\{b_1, b_2, \ldots\}^{(1)}$ obtained this way produces appreciable excitation error due to nonlinear terms in Eq. (11). Suppose that the difference in complex magnetization between the desired profile and the actual profile is $\Delta q_c(\vec{r}) = \Delta\left(\theta(\vec{r}) e^{i\phi_c(\vec{r})}\right)$. We make up for this difference by adding to the first solution a corrective RF pulse $\{b_1, b_2, \ldots\}^{(2)}$, which is obtained by solving Eq. (13) using $\Delta q_c(\vec{r})$ as a new target. If the procedure converges, the updated RF pulse $\{b_1, b_2, \ldots\}^{(1)} + \{b_1, b_2, \ldots\}^{(2)}$ should reduce the excitation error. The error should further diminish as higher order corrections are made in a similar manner.

Figure 3 illustrates the results of numerical simulation which demonstrate convergence of the procedure in two-dimensional excitation pulse design on an eight-channel parallel transmit system. The target excitation profile is a phase-coherent, 5 cm × 10 cm rectangle in the center of a 32 cm × 32 cm field of view. The normalized profile, shown in Fig 3(a), was scaled to three different tip angles 60°, 90°, 120° for separate simulations. Individual RF field maps $\{s_1, s_2, \ldots, s_8\}$ are shown in Fig. 4(c). For each tip angle, the first order waveforms $\{b_1, b_2, \ldots, b_8\}^{(1)}$ were determined by numerical inversion of Eq. (13) with an echo-planar gradient trajectory. The gradient pulse had six "blips" in the slow direction and the parallel transmit acceleration factor was $R = 2$. Off-resonance field was ignored. The inversion was done by a built-in algorithm (left division) of Matlab (Mathworks, Natick, MA, USA) with an appropriate RF power regularization parameter whose exact value did not affect our result significantly.

The Bloch-simulated complex excitation profiles $q_c(\vec{r})$ obtained from the first-order solutions are shown in the first row of Fig. 3(d); the lengths of the arrows are proportional to the tip angle $\theta$, and the directions indicate the transverse phase $\phi_c$. For tip angles of 90° (center column) and 120° (right column), deviation from the target profile (center column of Fig. 3a) is manifest. The second row of Fig. 3(d) shows how the profiles were corrected after the eighth iteration of the steps explained above. The reduction in the excitation error amplitude is shown in Fig. (3c). These maps display the amplitude of the complex difference between each of the maps in Fig. (3d) and the target profile Fig. (3a) (center). As we move from the initial solution ($N_{iter} = 1$) to the eighth ($N_{iter} = 8$), the maximum error amplitude is greatly reduced for all three tip angles.

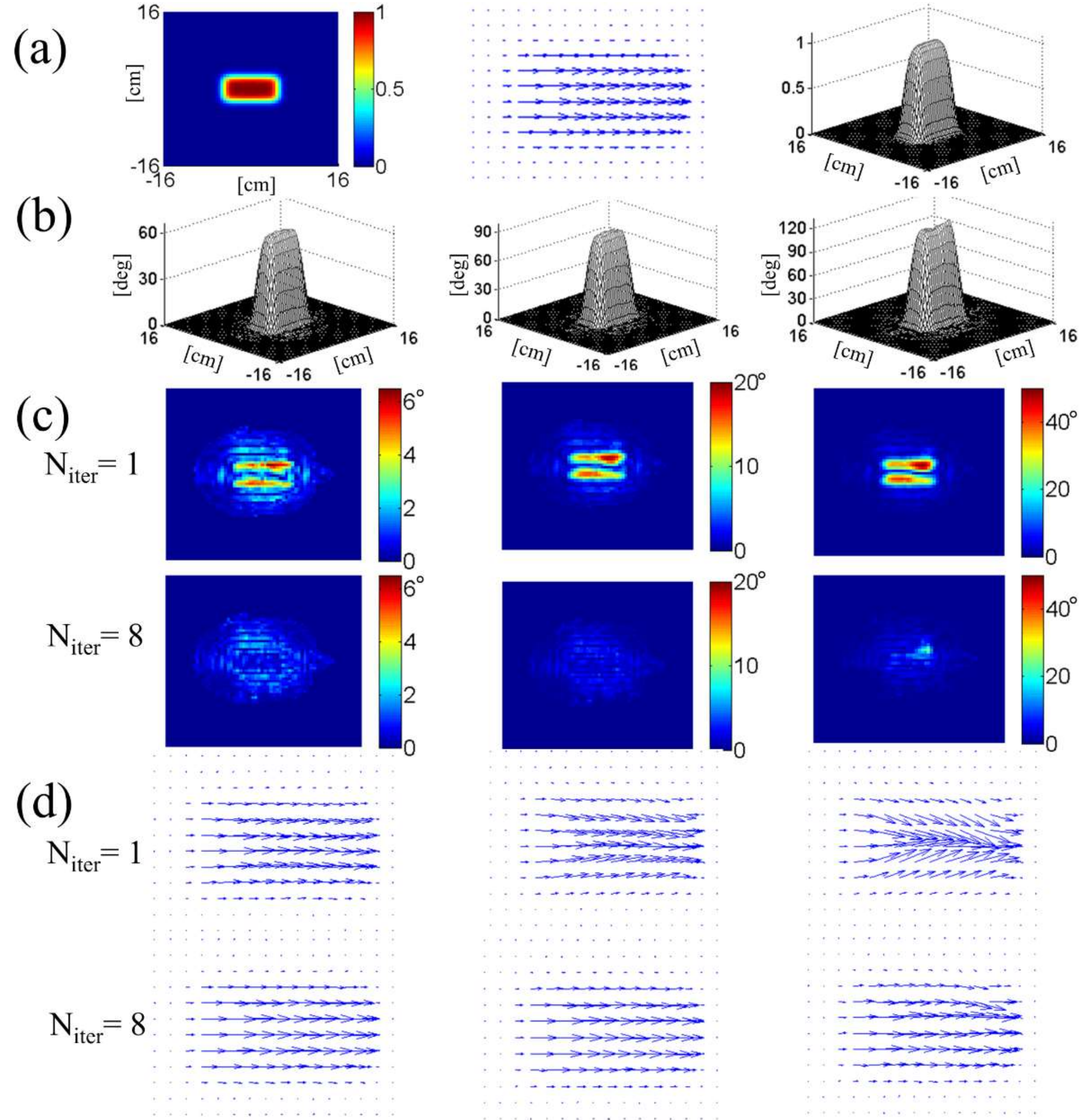

**Figure 3.** (a) Two-dimensional target excitation profile used in the numerical experiment on iterative inversion of the Bloch equation. Three different representations are shown, including a vector map of scaled magnetization $q_c$ (middle). The same target profile was scaled in magnitude for excitations with different tip angles. In (b-d), the left, middle, right columns correspond to tip angles 60°, 90°, 120°, respectively. (b) Excitation profiles after the 8-th order correction. (c) Excitation error amplitudes in degrees, obtained after Bloch-simulated magnetization $q_c(\vec{r}, T)$ at $N_{iter} = 1$ and 8 were subtracted from the target profile. (d) Complex excitation profiles at $N_{iter} = 1$ and 8, showing convergence of the phase and amplitude.

The progression of the added RF pulse energy and the maximum excitation error during the iteration process is displayed in Fig. 4. While both quantities quickly drop with $N_{iter}$ for tip angles 60° and 90°, at 120° the decrease is not as fast and is not monotonic; convergence is only marginal. This is expected because of rapid growth of nonlinear terms in Eq. (11) for $\theta > 90°$. This makes initial excitation error too big for simple iteration to correct, and necessitates more rigorous algorithm such as optimal control.

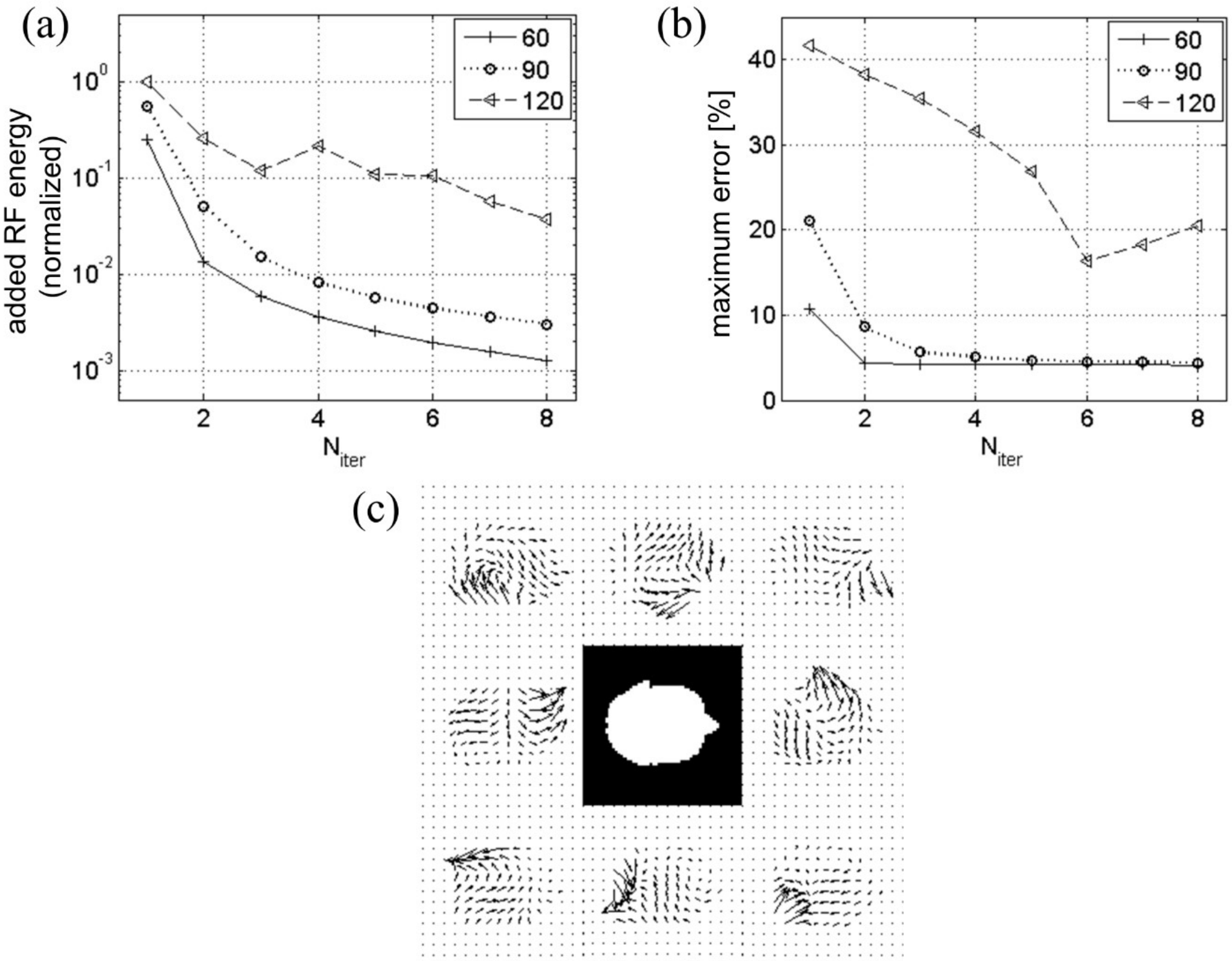

**Figure 4**. (a) Added RF energy from all coils at different iteration stages ($N_{iter}$) for tip angles 60°, 90°, 120°. Monotonic decrease in added RF energy for tip angles 60° and 90° indicates that iterative inversion converges for these angles. (b) Convergence is also indicated by the decrease in the excitation error measured as the maximum of the difference images (Fig. 3c) divided by the target tip angle. (c) Eight-coil $B_1^+$ maps used in the simulation. White mask in the center corresponds to an axial head slice over which the maps were calculated.

An important difference between our iterative method and the one in ref. [7] is that the present method enforces a predetermined excitation phase as well as the tip angle. This is because the method is based on the expansion of the full Bloch equation Eq. (11) expressed in the complex magnetization. In contrast, the additive angle method of ref. [7] aims at refining RF pulses for a desired tip angle profile only, creating phase-incoherent excitation. Although our method is limited to about 90° excitation angle, the fact that the resulting RF pulse excites phase-coherent magnetization means that larger tip-angle pulses could be constructed by RF concatenation. Luy et al [30] showed that if one can design a phase-coherent excitation pulse of tip angle $\theta$, one can construct an arbitrary $2\theta$ rotation pulse by adjoining the original pulse with its transformation. Extension of the method to multidimensional excitation and parallel transmit was demonstrated in [31].

### 4.2. Fast Bloch simulation

The local rotating frame approach can significantly reduce the Bloch simulation time if RF fields are much weaker than the gradient fields. In the new frame, the gradient fields are integrated out at the expense of additional rotation frequency imparted to the apparent RF field. The additional rotation can be much faster than conventional-frame RF nutation. However, since such rotation is exactly calculated and is not part of numerical simulation, numerical error resulting from finite integration step size grows much more slowly than when physical longitudinal field is present in the Bloch simulation. This point is illustrated by considering error accumulation in numerical integration of the differential equation Eq. (18),

$$\alpha_{n+1}^* = \alpha_n^* + \Delta t \frac{i\gamma}{2} B_{n+1}\beta_n$$

$$\beta_{n+1} = \beta_n + \Delta t \frac{i\gamma}{2} B_{n+1}^*\alpha_n^* \quad (35)$$

with initial condition $\alpha_0 = 1$, $\beta_0 = 0$ and $B_{n+1}, n = 0, 1, 2, \cdots$ is defined in Eq. (19).

Eq. (35) does not conserve the unitarity of the $(\alpha_n, \beta_n)$ pair. However, if the RF amplitudes are small enough, cumulative numerical error is small and does not affect the final magnetization significantly. In fact the norm of the Cayley-Klein parameters grows at each step of Eq. (35) as

$$|\alpha_{n+1}|^2 + |\beta_{n+1}|^2 = (|\alpha_n|^2 + |\beta_n|^2)\left(1 + \frac{\gamma^2\Delta t^2}{4}|B_{n+1}|^2\right). \quad (36)$$

After $N$ steps, the error becomes

$$|\alpha_N|^2 + |\beta_N|^2 - 1 \approx \frac{\gamma^2\Delta t^2}{4}\sum_{n=1}^{N}|B_n|^2 . \quad (37)$$

The right-hand side is one quarter of the sum of the square of incremental nutation angles (in radian), and has the order of magnitude of $N(\theta_t/N)^2 = \theta_t^2/N$. where $\theta_t$ is the total tip angle. For a 3.2 ms-long 90° pulse simulated with $\Delta t$ = 4 μs, $\theta_t^2/N = 0.0031 \ll 1$. In contrast, if the longitudinal field is not excluded from the simulation, the error grows as Eq. (36) but with $B_n$ replaced by the total magnetic field in the conventional rotating frame. The error can quickly run above unity at out-of-slice voxels where the longitudinal field is much larger than the RF field.

#### 4.2.1. Bloch simulation in parallel transmit

The following describes Bloch simulation with multiple transmit coils in the local rotating frame.

Step 1. (*phase table*) For a given off-resonance field $\delta B_0(r_m)$ and a gradient waveform $G_n = (G_{xn}, G_{yn}, G_{zn})$ on a time grid $t_n = n\Delta t$, $n = 1, 2, \cdots, N$, construct a spatio-temporal phase table $\phi_{mn} = \phi(r_m, t_n)$. Here $r_m = (x_m, y_m, z_m)$, $m = 1, 2, \cdots, L$ is the position of the *m*-th voxel, and $\phi_{mn}$ is calculated from Eq. (1). This step needs to be done only once if gradient pulse is predetermined, or every time gradient pulse is changed.

Step 2. (*frame transformation*) The RF waveforms in conventional rotating frame of all coils are combined with the respective $B_1^+$ maps as in Eq. (34) to determine complex RF fields at each time and voxel point. This gives an $L$ by $N$ matrix representing RF fields in the conventional frame. The RF field in the local rotating frame is then obtained by element-by-element multiplication of this matrix with $\exp(i\phi_{mn})$, as in Eq. (4).

Step 3. (*integration*) In the spinor domain, solve recursively the difference equation Eq. (35) to obtain $\alpha_N, \beta_N$ at the end of the pulse.

Step 4. (*back transformation*) The Cayley-Klein parameters are transformed back to the conventional rotating frame. For an inherently-refocused gradient pulse and negligible off-resonance field, however, this step can be omitted since $\phi_{mN} = 0$ for every voxel. Otherwise, the back transformation is done by multiplying $\alpha_N(r_m)$, $\beta_N(r_m)$ with $\exp\,(i\phi_{mN}/2)$ according to Eq. (16).

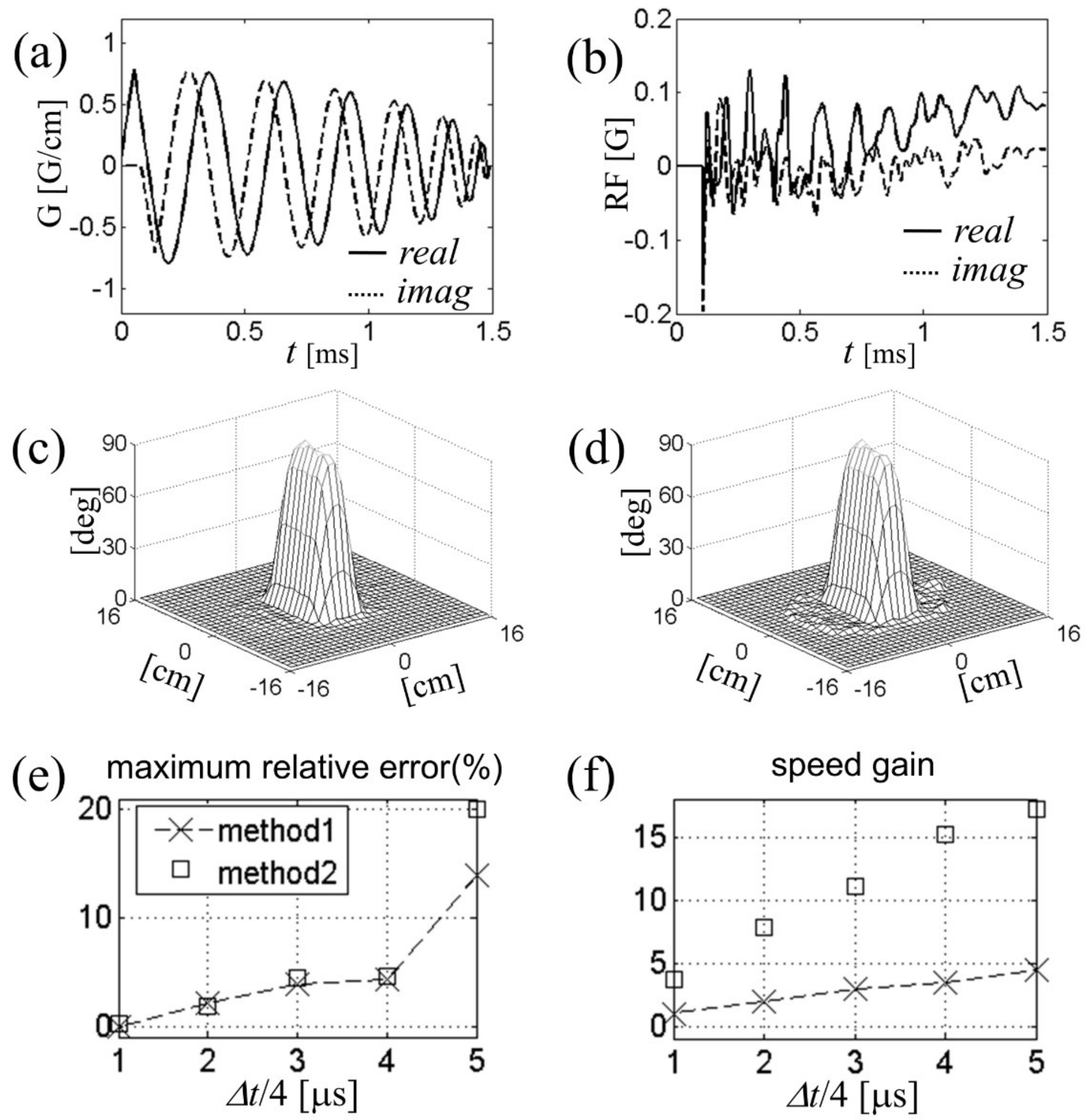

**Figure 5**. (a) Six-turn spiral gradient waveform expressed as $G = G_x + iG_y$ used in the eight-channel parallel excitation simulation. (b) RF field in the conventional frame at the center of the field of view. (c-d) Tip angle profiles for rectangular region-of-interest excitation. Bloch simulation was performed in two methods defined in the text. (e-f) Accuracy and speed of Bloch simulation as a function of the step size $dt$.

In Fig 5 we compare the speed and accuracy of this method (called Method 2) with those of conventional spinor-domain Bloch simulation in which Cayley-Klein parameters advance on the same time grid through explicit complex rotation in the conventional frame (Method 1). Five different time steps, $\Delta t/4$ = 1, 2, ..., 5 μs were used in separate simulations. For concreteness we again took an example of two-dimensional reduced region-of-interest excitation with $B_1^+$ maps of Fig. (4c). The test RF waveforms were Fourier-designed on a spiral gradient trajectory (Fig. 5a) for 90° excitation of a rectangular region. Typical eight-coil combined RF field is shown in Fig. (5b). All calculations were done with Matlab on a personal computer with 2 gigabytes of memory and 1.8 GHz dual-core CPU.

Figure (5e,f) show the relative simulation error and speed gain as a function of the step size for both methods. For both error and speed calculations, we used Method 1 with $\Delta t$ = 4 μs as a "reference" simulation; it is considered exact if RF and gradient pulses are actually digitally updated every 4 μs. Each data point in Fig. (5e) corresponds

to the maximum tip angle error divided by 90° at the end of the simulation, whereas Fig (5f) shows the inverse simulation time normalized to that of the reference simulation. We find that up to $\Delta t$ =16 μs, Method 2 matches the accuracy of Method 1, both generating less than 5% of error in the simulation result. For a given $\Delta t$, on the other hand, Method 2 is faster than Method 1 by a factor of about 4. The tip angle profile for Method 2, $\Delta t$ =16 μs is displayed in Fig. (5d). The profile shows little difference from the reference simulation (Fig. 5c).

**4.2.2. Application to optimal control**

Fast Bloch simulation can directly benefit the optimization speed in parallel-transmit RF pulse design based on the optimal control theory [8], [32], [33], [34]. The optimal control design starts with an initial RF pulse typically obtained by Fourier design, and iteratively updates it through the following steps: calculate optimization direction in the possible RF pulse space → line search to minimize the cost function → update RF pulse. The optimization direction is found through two Bloch simulations, forward and backward [35], and the line search utilizes several forward Bloch simulations for cost evaluation.

Figure 6 shows an example of optimal control RF design in which the Bloch simulations for line search were expedited by local rotating frame transformation. The initial RF pulse, gradient trajectory, and the $B_1^+$ maps were taken from the previous simulation (Fig. 5), except that the RF amplitude was doubled for 180° excitation. Figure (6a) shows the initial tip angle profile with significant error. The cost function [8] was the sum of the excitation error, defined as the sum of squares of the deviation in the Cayley-Klein parameters, and the total RF power multiplied by a regularization parameter. Figure (6b) shows the optimized tip angle profile obtained after twenty updates. Each update involved a line search with ten Bloch simulations in the local rotating frame. The square marker trace in Fig (6c) shows the decrease in the cost function as a function of the CPU time. In comparison with the same RF calculation with conventional spinor-domain Bloch simulation (Method 1), about 2.5-fold improvement in the optimization speed is observed. Further improvement was limited by calculation of the cost function and its gradient, for which no acceleration was attempted.

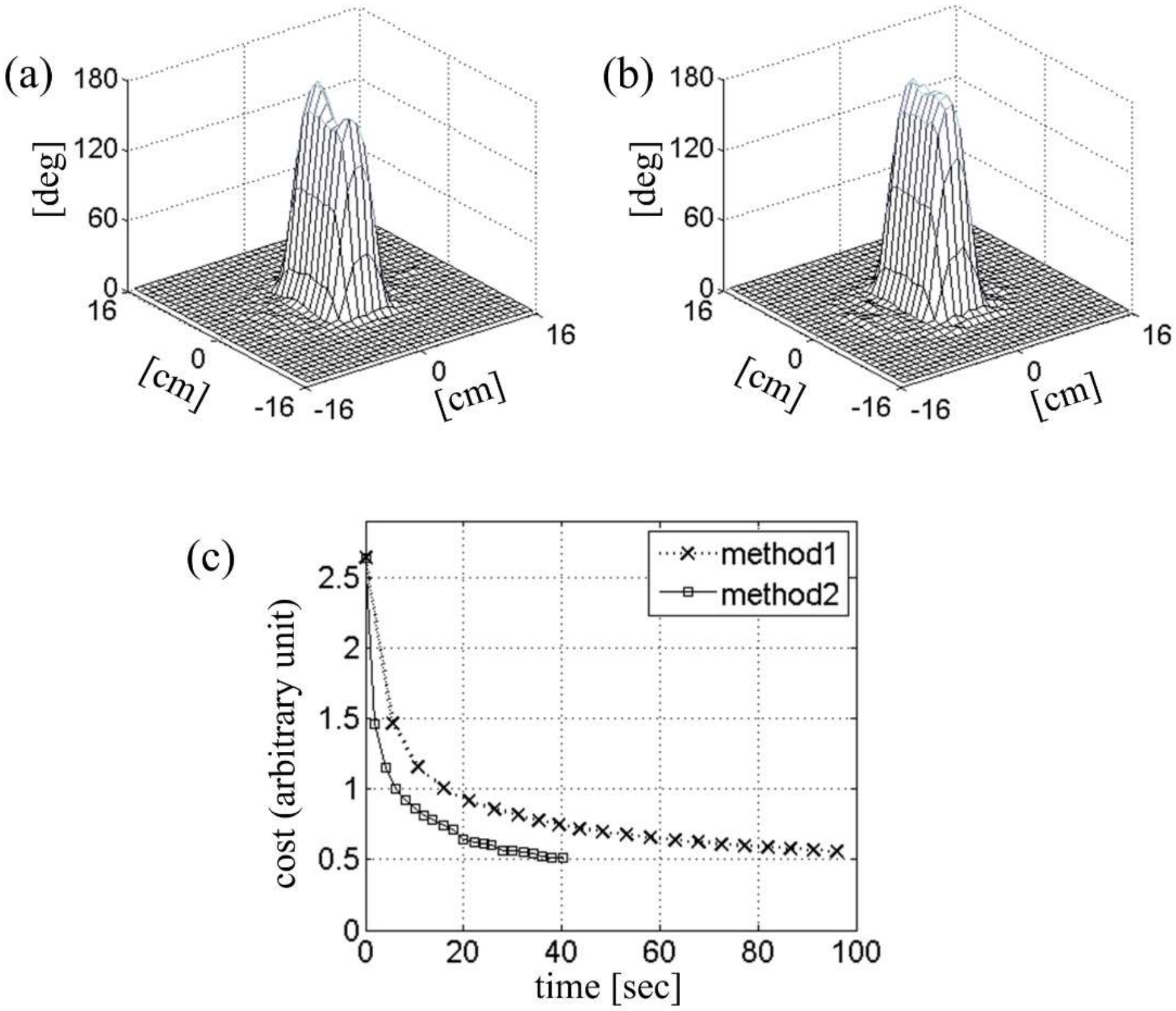

**Figure 6**. Two-dimensional inversion profiles obtained (a) before and (b) after optimal-control iterative correction. (c) Cost as a function of time during optimization. The same two Bloch simulation methods from Fig. (5c-d) were utilized in line-search subroutines.

## 5. Summary and Discussion

In this paper we presented a general theoretical framework in which spatially selective RF pulses are calculated in a longitudinal-field-free frame of reference. The method is conceptually straightforward, and is relevant to situations in which RF pulses are optimized for a predetermined gradient pulse. Given relative scarcity of analytical methods in multi-coil RF design, one motivation for the present theoretical work has been to streamline the numerical workflow and speed up calculations in iterative RF design. As shown in section 4, we found that the new approach provides significant benefit in accurate estimation of a 90° pulse and nonlinear optimization of a large tip angle RF pulse for multiple coils.

When applied to a single-coil slice selective excitation, we found that the local rotating frame approach and change of the magnetization variable provide theoretical insights which have not been necessarily obvious. As examples, we presented alternative derivation of the forward Shinnar-Le Roux transformation and explicit calculation of the nonlinear phase in the slice selective excitation. The tip angle dependence of the nonlinear phase (Eq. (33)) should be of practical use in determining the optimum area of the rewinder gradient following slice selective excitation for a range of tip angles.

An interesting theoretical observation from the present work is that when the gradient fields are transformed away, the Bloch equation in the resulting RF-only frame can be well approximated by a linear response equation for tip angles much larger than 30°. This can be seen from Eq. (11) as well as from Eq. (27) in which the small parameter for linear approximation is one half of the tipping angle. This observation correctly suggests that linear design based on $\theta e^{i\phi_c}$ should be valid for intermediate tip angles, even if conditions for an LCLTA pulse are not satisfied. In section 4, we showed that this is the case to the extent that errors due to linear approximation can be readily corrected by iterative inversion. This provides a fast method to obtain a high-quality, phase-restricted 90° pulse in multidimensional excitation. The fact that the designed pulse is phase specific also suggests that such pulses can be concatenated for construction of a 180° pulse.

We note that several authors have published on novel approaches to solve or interpret the Bloch equation, including spherical coordinate description [36], [37], hybrid state equations [38], and novel variable transformation to reduce nonlinearity [39]. In particular, Boulant and Hoult [39] laid out a theory to extend the small-tip-angle regime through a complex tip angle-based variable transformation, focusing on applications to slice-selective excitation.

A promising direct application of the local rotating frame transformation is acceleration of Bloch simulation in the presence of strong gradient fields. For optimization of a relatively long spatial RF pulse with a high-resolution spatial profile, the benefit of fast calculation of the cost function could be substantial. Since the new frame shares the same longitudinal axis as the laboratory frame, $T_1$ and $T_2$ relaxation can also be incorporated into the simulation in a straightforward manner. Any speed gain through the frame transformation can also be augmented by independent acceleration from computing hardware enhancement such as with graphics processing units [40].

High quality initial estimate for a large-tip-angle pulse and fast optimization algorithm are both important for patient-specific RF pulse design in high-field MRI. We envision that the methods presented here will help develop practical solutions for RF pulse optimization based on patient-dependent $B_1^+$ maps generated on the fly in high field. Evaluating the feasibility and robustness of such solutions in more specific situations will be a subject of future research.

**Appendix. Residual phase winding in slice selective excitation**

First we show the relationship between $\alpha^{(2)}(T)$ and $|\beta^{(1)}(T)|$. The third line of Eq. (27) can be rewritten as

$$\alpha^{(2)}(T) = -\frac{\gamma^2}{4}\int\int_{0<t'<t<T} B^*(\vec{r},t)B(\vec{r},t')dt'dt\,. \qquad (A1)$$

Conjugating the right-hand side is equivalent to swapping $t$ and $t'$ inside the integral. Such swapping in turn is equivalent to changing the integral domain from $0 < t' < t < T$ to $0 < t < t' < T$. The new domain complements the original domain to cover the two-dimensional square domain $0 < (t, t') < T$. Therefore the sum of $\alpha^{(2)}(T)$ and $\alpha^{*(2)}(T)$ is a separable integral

$$\alpha^{(2)}(T) + \alpha^{*(2)}(T) = 2\alpha^{(2)}_{\text{real}}(T) = -\frac{\gamma^2}{4}\int_0^T B^*(\vec{r},t)dt\int_0^T B(\vec{r},t')dt' = -\left|\beta^{(1)}(T)\right|^2. \qquad (A2)$$

This is our first relationship. Next, we substitute $B(\vec{r},t) = B_c(t)e^{i\gamma Gzt}$, $0 < t < T$, in the integral in Eq. (A1), and change the variables $(t, t') \rightarrow (s \equiv t - t', t')$. This converts the integral into a Fourier transform on the positive half of the real axis

$$\alpha^{(2)}(z,T) = \int_0^\infty F(s) e^{-i\gamma G z s} ds \tag{A3}$$

where we explicitly indicated the $z$-dependence of $\alpha^{(2)}$ and

$$F(s) \equiv -\frac{\gamma^2}{4} \int_0^{T-s} B_c^*(s+t') B_c(t') dt' \,. \tag{A4}$$

In the above we also defined without loss of generality $B_c(t) = 0$ for $t < 0$ and $t > T$. The real and imaginary parts of a Fourier integral of an analytical function on the positive half axis satisfy the dispersion relation [41],

$$\alpha_{\text{imag}}^{(2)}(z,T) = -\frac{1}{\pi} \int_{-\infty}^{\infty} \frac{\alpha_{\text{real}}^{(2)}(z',T)}{z-z'} dz' \tag{A5}$$

which is the second relationship we need. Eqs (A2, A5) show that $\alpha^{(2)}(z,T)$ is completely determined by the first-order excitation profile $\left|\beta^{(1)}(z,T)\right|$.

Next, we outline the omitted steps in Eq. (33). All the quantities below are evaluated at $t = T$. Here we only indicate the $z$-dependence of $\alpha$ and $\beta$. We assume a symmetric slice profile, $\left|\beta^{(1)}(-z)\right| = \left|\beta^{(1)}(z)\right|$. Then the calculation of the second line in Eq. (33) proceeds as follows. From Eq. (A2),

$$-\frac{d}{dz}\left(\frac{\alpha_{\text{imag}}^{(2)}}{1+\alpha_{\text{real}}^{(2)}}\right)\Bigg|_{z=0} = -\frac{d}{dz}\left(\frac{\alpha_{\text{imag}}^{(2)}}{1-\left|\beta^{(1)}\right|^2/2}\right)\Bigg|_{z=0} . \tag{A6}$$

Since the denominator in the differential is an even function of $z$, we can take it out as

$$-\frac{d}{dz}\left(\frac{\alpha_{\text{imag}}^{(2)}}{1-\left|\beta^{(1)}\right|^2/2}\right)\Bigg|_{z=0} = -\frac{1}{1-|\beta^{(1)}(0)|^2/2} \cdot \frac{d\alpha_{\text{imag}}^{(2)}}{dz}\Bigg|_{z=0} . \tag{A7}$$

From Eq. (A5), the derivative on the right-hand side is

$$\begin{aligned} \frac{d\alpha_{\text{imag}}^{(2)}}{dz} &= -\frac{1}{\pi}\int_{-\infty}^{\infty} \frac{d}{dz}\left(\frac{1}{z-z'}\right) \cdot \alpha_{\text{real}}^{(2)}(z') dz' \\ &= \frac{1}{\pi}\int_{-\infty}^{\infty} \frac{d}{dz'}\left(\frac{1}{z-z'}\right) \cdot \alpha_{\text{real}}^{(2)}(z') dz' \\ &= -\frac{1}{\pi}\int_{-\infty}^{\infty} \frac{d\alpha_{\text{real}}^{(2)}(z')}{dz'} \cdot \frac{1}{z-z'} dz' \,. \end{aligned} \tag{A8}$$

Substituting $\alpha_{\text{real}}^{(2)}(z') = -\left|\beta^{(1)}(z')\right|^2/2$ and $z = 0$, we obtain

$$\frac{d\alpha_{\text{imag}}^{(2)}}{dz}\Bigg|_{z=0} = -\frac{1}{2\pi}\int_{-\infty}^{\infty} \frac{1}{z'}\frac{d}{dz'}\left|\beta^{(1)}(z')\right|^2 dz' = -\frac{1}{\pi}\int_0^{\infty} \frac{1}{z'}\frac{d}{dz'}\left|\beta^{(1)}(z')\right|^2 dz' \,. \tag{A9}$$

The last equation holds because the integrand is an even function. For any top-hat-like profile, the integrand is peaked near the slice boundary. This motivates us to define an effective slice thickness $z_{\text{FW,eff}}$ as

$$\frac{1}{z_{\text{FW,eff}}/2} \equiv \frac{\int_0^\infty \frac{1}{z'}\frac{d}{dz'}\left|\beta^{(1)}(z')\right|^2 dz'}{\int_0^\infty \frac{d}{dz'}\left|\beta^{(1)}(z')\right|^2 dz'} = -\frac{1}{\left|\beta^{(1)}(0)\right|^2}\int_0^\infty \frac{1}{z'}\frac{d}{dz'}\left|\beta^{(1)}(z')\right|^2 dz' \tag{A10}$$

which then reduces Eq. (A9) to

$$\frac{d\alpha_{\text{imag}}^{(2)}}{dz}\Bigg|_{z=0} = \frac{2}{\pi}\frac{\left|\beta^{(1)}(0)\right|^2}{z_{\text{FW,eff}}} . \tag{A11}$$

If we substitute this result to the right-hand side of Eq. (A7), and notice that $\left|\beta^{(1)}(0)\right| = \theta_t/2$ is the half tip-angle, we finally obtain

$$\left.\frac{d\phi_{res}}{dz}\right|_{z=0} = -\frac{1}{2\pi z_{\mathrm{FW,eff}}} \cdot \frac{\theta_t^2}{1-\theta_t^2/8}. \qquad (A12)$$

The residual phase slope is negative, and varies quadratically with the tip angle for small tip angles. Under our convention, negative residual slope means that more negative gradient field is needed to completely "rewind" the transverse phase, which is consistent with experimental findings.

## Acknowledgements

The author acknowledges helpful discussions with Drs. Dan Xu, Ileana Hancu, and Mika Vogel. This work was partly supported by NIH grant 5R01EB005307-2. The content is solely the responsibility of the author and does not necessarily represent the official views of the NIH.